\documentclass[aps,prb,10pt,showpacs,superscriptaddress,twocolumn,longbibliography]{revtex4-1}
\usepackage{graphicx}
\usepackage{amssymb}
\usepackage{amsmath}
\usepackage{color}
\usepackage{paralist}
\usepackage{bm}
\usepackage{verbatim}

\def\cH{\hat{\cal H}}
\def\hh{{\hat{h}}}

\def\bB{{\bf B}}
\def\bd{{\bf d}}

\def\bk{{\bf k}}
\def\bK{{\bf K}}

\def\bq{{\bf q}}
\def\bJ{{\bf J}}
\def\bp{{\bf p}}
\def\br{{\bf r}}

\def\hbsigma{\hat{\boldsymbol \sigma}}
\def\hbnabla{\hat{\boldsymbol \nabla}}

\def\bomega{{\boldsymbol \omega}}

\def\ha{\hat a}
\def\hb{\hat b}
\def\hd{\hat d}

\def\hh{\hat h}

\def\hV{\hat V}

\def\hsigma{\hat\sigma}

\def\ty{\mathrm{y}}
\def\tj{\mathrm{j}}

\newcommand{\ket}[1]{| #1 \rangle}

\begin{document}

\title{Emergent Weyl excitations in systems of polar particles}

\author{Sergey V. Syzranov }
\email{sergey.syzranov@googlemail.com}
\affiliation{Physics Department, University of Colorado, Boulder, CO 80309}
\affiliation{JILA, NIST, University of Colorado, Boulder, CO 80309}
\affiliation{Center for Theory of Quantum Matter, University of Colorado, Boulder, CO 80309}

\author{Michael L. Wall }
\affiliation{JILA, NIST, University of Colorado, Boulder, CO 80309}
\affiliation{Center for Theory of Quantum Matter, University of Colorado, Boulder, CO 80309}

\author{Bihui Zhu}

\affiliation{Physics Department, University of Colorado, Boulder, CO 80309}
\affiliation{JILA, NIST, University of Colorado, Boulder, CO 80309}
\affiliation{Center for Theory of Quantum Matter, University of Colorado, Boulder, CO 80309}


\author{Victor  Gurarie}
\affiliation{Physics Department, University of Colorado, Boulder, CO 80309}
\affiliation{JILA, NIST, University of Colorado, Boulder, CO 80309}
\affiliation{Center for Theory of Quantum Matter, University of Colorado, Boulder, CO 80309}

\author{Ana Maria Rey}
\email{arey@jilau1.colorado.edu}
\affiliation{Physics Department, University of Colorado, Boulder, CO 80309}
\affiliation{JILA, NIST, University of Colorado, Boulder, CO 80309}
\affiliation{Center for Theory of Quantum Matter, University of Colorado, Boulder, CO 80309}

\date{\today}

\begin{abstract}
  Weyl fermions are  massless chiral particles first  predicted  in 1929 and once thought to describe neutrinos.
Although never observed as elementary particles, quasiparticles with Weyl dispersion have recently been experimentally discovered in solid-state systems  causing a furore in the research community. Systems with Weyl excitations can display a plethora of fascinating phenomena and offer great potential for improved quantum technologies. Here we show that Weyl excitations generically exist in three-dimensional systems of dipolar particles with weakly broken time-reversal symmetry (for example, by a magnetic field). They emerge
as a result of dipolar-interaction-induced transfer of angular momentum between the $J=0$ and $J=1$ internal particle levels.
We also discuss momentum-resolved Ramsey spectroscopy methods for observing Weyl quasiparticles in cold alkaline-earth-atom systems. Our results provide a pathway for a feasible  experimental realisation  of Weyl quasiparticles and related phenomena in clean and controllable atomic systems.
\end{abstract}

\maketitle

\section{Introduction}
Recent predictions\cite{Wan:WeylProp,BalentsBurkov} and experimental
observations\cite{Hasan:DiscoveryArcs,Weng:PhotCrystWSM,Lu:PhotCrystWSM} of
Weyl semimetals in solid state systems have instigated
intensive studies of their properties, such as  non-local electrodynamics and chiral anomaly \cite{Parameswaran:Anomaly},
topologically protected Fermi arcs on the surfaces \cite{Wan:WeylProp,Hasan:DiscoveryArcs,Weng:PhotCrystWSM}, non-Anderson disorder-driven transitions\cite{Fradkin2,GoswamiChakravarti,Syzranov:Weyl,Syzranov:unconv,Brouwer:WSMcond,Moon:RG,Herbut} and  unusual dependencies
of transport properties on doping and temperature\cite{Hwang:WSMcond,Skinner:MinCond,Rodionov:Coulomb}. In parallel, enormous research efforts are now directed at
finding Weyl excitations in new systems. A promising platform for exploring Weyl physics
is tunable and fully controllable ultracold atomic gases\cite{Lan:WSMprop,Jiang:WSMprop,Spielman:WSMprop,Soljacic:WSMprop,Ganeshan:WSMprop,Liu:SFimplementation}.
However, henceforth proposed cold-atom realisations of Weyl quasiparticles have focussed mostly on non-interacting systems, and all have required implementations of externally imposed spin-orbit coupling through laser assisted tunneling schemes\cite{Lan:WSMprop,Jiang:WSMprop,Spielman:WSMprop,Soljacic:WSMprop,Ganeshan:WSMprop}, other optical means~\cite{1674-1056-24-5-050502,gong2011bcs,seo2012emergence}, or external rotating fields\cite{Liu:SFimplementation}.

In this paper we demonstrate that excitations with Weyl dispersion generically emerge
in three-dimensional (3D) arrays of dipolar particles in the presence of a weak magnetic field,
as a result of the dipole-interactions-induced transitions between their internal angular-momentum $J=0$ and $J=1$ states.
These excitations exhibit the same single-particle physics as Weyl fermions \cite{Weyl:first}  but, similarly to other non-fermionic Weyl excitations~\cite{Lu:PhotCrystWSM}, their many-particle  properties are expected to be different, opening alternative research directions, new functionalities and applications beyond those accessible with   solid-state systems~\cite{Hasan:DiscoveryArcs,Weng:PhotCrystWSM}.

We also show that experimentally such excitations can be observed, for instance, in trapped alkaline-earth atoms (AEAs)
in a 3D optical lattice with lattice spacings smaller than the wavelength of the electronic $J=0-J=1$ transition.
The  simple and unique internal structure of these atoms has already lead to record levels of precision and accuracy in atomic clocks\cite{Bloom2014}.  Taking advantage of the well developed spectroscopic tools to interrogate and manipulate AEAs, we propose to  probe the Weyl quasiparticle dispersion and non-trivial chirality  by means of   momentum-resolved Ramsey  spectroscopy. Our proposal opens a path  for a feasible  experimental realisation  of Weyl quasiparticles in clean and controllable atomic systems. Moreover, it lays the groundwork for the yet unexplored regime of topologically protected sensing, owing to the topological robustness of Weyl quasiparticles that could be used to push the stability and accuracy of optical-lattice AEA-based clocks.


\section{Results}

{\bf Phenomenological argument.}
We assume that the system has long-lived excitations (quasiparticles) with (integer) angular momentum $\bJ$.
Due to the translational invariance, the  (quasi)momentum $\bk$ is a good
quantum number. In the long-wave limit the effective quasiparticle Hamiltonian is insensitive to
the details of the potential of the periodic lattice that the particles may be placed in.
To preserve rotation and inversion symmetries in the absence of magnetic field the Hamiltonian has
to be an even function of $(\bk \cdot \hat\bJ)$ and a function of $|\bk|$ and $\hat\bJ^2$.
In the presence of a sufficiently weak uniform magnetic field,   $\bomega$, that creates a perturbation $-\bomega\cdot\hat\bJ$ independent of
$\bk$ in the limit $\bk\rightarrow0$, the most generic form of the quasiparticle Hamiltonian is given by
\begin{equation}
	\hh(\bk,\bJ)=F\left[|\bk|,(\bk \cdot \hat\bJ)^2,\hat\bJ^2\right]-\bomega\cdot\hat\bJ,
	\label{HamInit}
\end{equation}
where $F$ is an arbitrary function of three arguments.

The small quasimomentum $\bk$ can be measured from any high-symmetry point
in the Brillouin zone characterised by an isotropic dispersion $\xi_{\bk}=\xi(|\bk|)$ of
non-interacting particles in the limit $\bk\rightarrow0$.

For the particular case of $J=1$, the Hamiltonian~\eqref{HamInit} has nodes at momenta $\bK \parallel\bomega$, such that $F(|\bK|,|\bK|^2,2)\pm\omega
=F(|\bK|,0,2)$, corresponding to two intersecting branches with angular momentum projections $J_{\rm z}=0$ and $J_{\rm z}=1$
or $J_{\rm z}=-1$ on the direction of magnetic field.
We note that such nodes always exist for weak magnetic fields and Hamiltonians that are regular as a function
of $\bk$.

 The excitation Hamiltonian near a node is obtained by
expanding the function $F$ in small momentum $\bp=\bk-\bK$. For a 3D system, it has Weyl dispersion of the form (see Methods):
\begin{align}
	\hh_{\text{eff}}(\bp)=\zeta(\bp)+v_\bot\hat\sigma_{\mathrm{x}} p_{\mathrm{x}}+v_\bot\hat\sigma_{\mathrm{y}} p_{\mathrm{y}}+v_\parallel\hat\sigma_{\mathrm{z}} p_{\mathrm{z}}
	\label{WeylDisp}
\end{align}
with Pauli matrices $\hsigma_i$ acting in the space of the respective two angular-momentum projections.

{\bf Model.} In what follows we confirm the above phenomenological argument by microscopic calculations
for a 3D system of dipolar particles described by the Hamiltonian
\begin{align}
\hat{H}&=\sum_{i}\hat{H}_0\left(\mathbf{r}_i\right)+\frac{1}{2}\sum_{i,j}\hat{H}_{\mathrm{dip}}\left(\hat{\mathbf{d}}_i,\hat{\mathbf{d}}_j,\mathbf{r}_i-\mathbf{r}_j\right)\, ,
\end{align}
where $\hat{\mathbf{d}}_i$ is the dipole moment operator of the $i$-th particle, and
\begin{equation}
\hat{H}_0\left(\mathbf{r}_i\right)=-\frac{\hbnabla_i^2}{2m}+U(\br_i)+ B_{\mathrm{J}} \hat\bJ_i^2-\hat\bJ_i\cdot \bB
\end{equation}
is the single-particle Hamiltonian that
includes the particle kinetic energy $-\frac{\hbnabla_i^2}{2m}$ (hereinafter $\hbar=1$),
the periodic potential $U(\br_i)$ of the lattice that the system may be placed in,
the  energy $ B_{\mathrm{J}}\hat{\bJ}_i^2$ of internal levels with $\hat\bJ_i$ being the angular momentum of the $i$-th particle,
 and the interaction $-\hat\bJ_i \cdot \bB$ with magnetic field (measured in units of the gyromagnetic ratio) that splits the $J=1$ levels.

The most generic form of the dipole-dipole interaction, which accounts for retardation effects, is given by \cite{GrossHaroche}(see also Methods)
\begin{align}
	\hat{H}_{\mathrm{dip}}(\hat\bd_i,\hat\bd_j,\br)
	=a(r)(\hat{\bf {r}}\cdot  \hat\bd_i)(\hat{\bf {r}} \cdot \hat\bd_j)+b(r)\hat\bd_i\cdot \hat\bd_j
	\label{DipoleInteraction}
\end{align} where $\hat{\bf {r}}={\bf r}/r$; $a(r)= \frac{3\gamma_0}{4d^2}[\ty_2(k_0 r)-i \tj_2(k_0 r)]$ and $b(r)=\frac{3\gamma_0}{4d^2}\sum_{n=0}^{1}(-1)^n [\ty_n(k_0 r)-i \tj_n(k_0 r)]/(k_0 r)$
{for $r\neq0$,}
with $\ty_n$  and $\tj_n $ being the $n$-th-order
spherical Bessel functions of the second and first  kind respectively and  $k_0$-- the wavevector of the $J=0$ to $J=1$ transition. The terms proportional to $\ty_n$ describe elastic interactions between dipoles a distance $r$ apart, while the terms with ${\rm{j}}_n$ account for the inelastic collective photon emission (radiation). $\gamma_0=\frac{k_0^3  d^2}{3 \pi \epsilon_0}$ is the natural linewidth of the transition and $d$ is {its} dipole moment.
If the dipoles are much closer to one another than the wavelength of the dipole transition, $k_0 r\ll 1$, retardation effects can be ignored, and one recovers the more familiar form of the dipolar interactions,
$a(r)\propto -3/r^3$, $b(r)\propto 1/r^3$,
common for
NMR solid-state systems\cite{Childress:diamond}, polar molecules\cite{Yan2013} and Rydberg atoms\cite{Ott:Rydberg,Robicheaux:Rydberg}.

We note that the above phenomenological derivation of the dispersion of Weyl-type quasiparticles
carries over straightforwardly to other dimensions. For example, a 2D system of dipolar particles with an in-plane
magnetic field hosts 2D Dirac excitations with the dispersion of monolayer graphene \cite{Novoselov666}. We emphasise that such 2D excitations
are distinct from the 2D ``chiron'' excitations\cite{Syzranov:chirons} that exist in a perpendicular magnetic field and resemble electrons
in bilayer graphene.

\begin{figure*}[b]
	\centering
	\includegraphics[width=.8\textwidth]{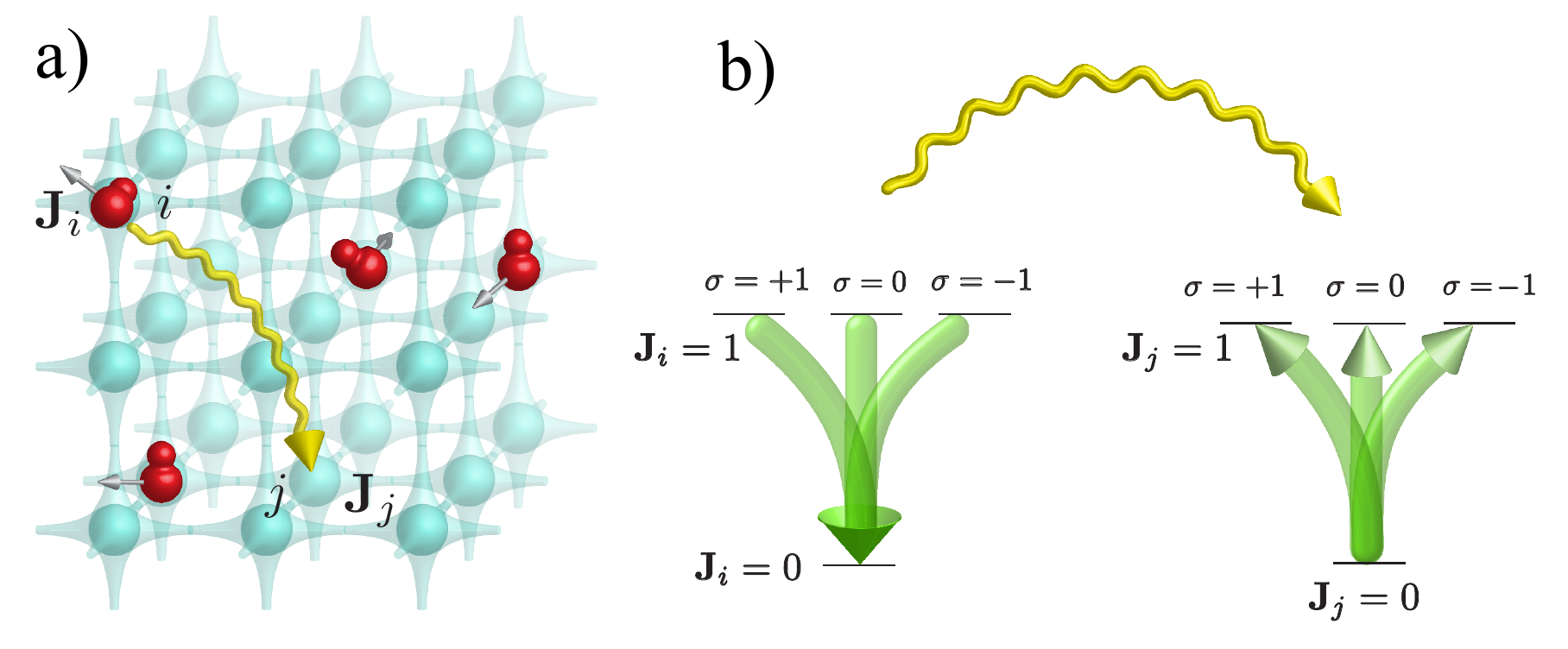}
	\caption{ {\bf Weyl quasiparticles in 3D dipolar arrays} (a) Schematics of the 3D lattice potential that traps an array of dipolar particles. The lattice  is deep enough to pin the particles, most of which  are prepared in the $J=0$ ground state (blue spheres). Only a few particles are excited to the $J=1$ states. Dipolar interactions between the $J=0$ and $J=1$ states  give raise to Weyl excitations.  (b) Schematics of dipole mediated interactions: An excited $J = 1$ state of one particle can be transferred to another particle in the $J = 0$ state by dipole-dipole interactions (virtual photon exchange is shown with a yellow wiggly line). Three types of allowed processes include $\ket{00}_i\ket{1\sigma}_j\leftrightarrow\ket{1\sigma}_i\ket{00}_j$,
$\ket{00}_i\ket{10}_j\leftrightarrow\ket{1,\pm1}_i\ket{00}_j$, and
$\ket{00}_i\ket{1,\pm1}_j\leftrightarrow\ket{1,\mp1}_i\ket{00}_j$.}
	\label{Lattice}
\end{figure*}

{\bf Atoms in a deep lattice.}
{While the above phenomenological argument demonstrates the existence of Weyl quasiparticles in a generic
3D system of dipolar particles in magnetic field, below we focus on the experimentally important case
of particles pinned in a deep unit-filled cubic lattice (Fig.~\ref{Lattice}a)
with small lattice spacing $a$; $a k_0\ll 1$.}

{
We assume that all particles are initially prepared in the $J_i=0$ state and that the
energy $B_{\mathrm{J}}$ of internal levels significantly exceeds the interaction strength
(usually in dipolar gases\cite{Yan2013,ShreckLesanovsky} $|\hat{H}_{\mathrm{dip}}|/B_{\mathrm{J}}\lesssim 10^{-6}$),
leading to the conservation of the number of sites excited to the $J=1$ state to a good
accuracy (cf. Methods).}

\begin{figure*}[t]
	\centering
	\includegraphics[width=\textwidth]{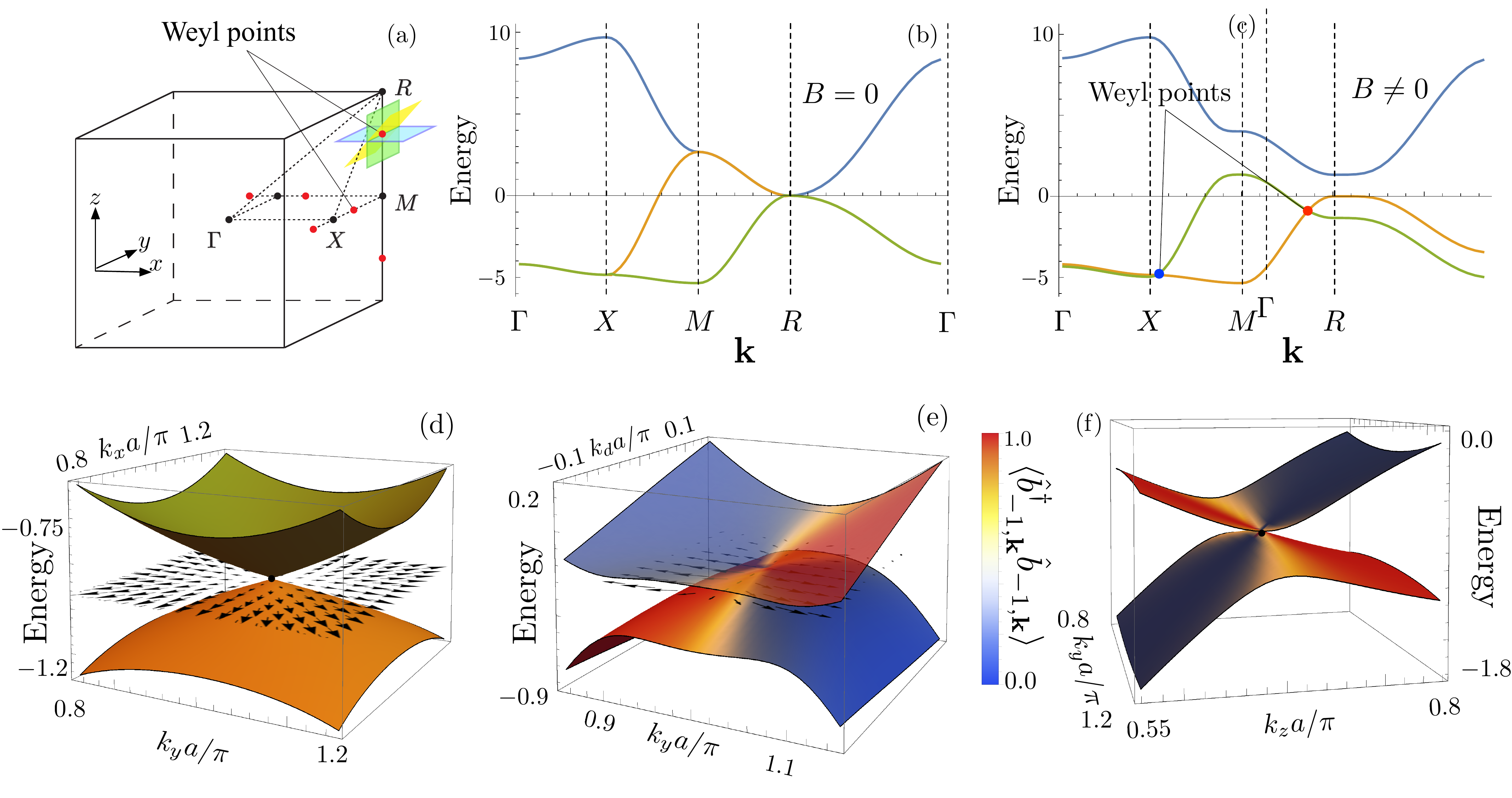}
	\caption{ {\bf Weyl quasi-particle dispersion and eigenstates} (a) Brillouin zone for the simple cubic lattice.
	(b) Dispersion along high-symmetry lines in the absence of magnetic field (all energies are measured in units of $(3/4)\gamma_0/(k_0a)^3$).
	(c) Dispersion in the presence of magnetic field $B=\gamma_0/(k_0a)^3$
	demonstrating the existence of Weyl nodes (red points) with linear quasiparticle dispersion near them.
	Each node is characterised by the monopole charge $\pm1$.
	In agreement with the fermion doubling theorem\cite{NielsenNinomiya} (the Nielsen-Ninomiya no-go theorem),
	there is an even number (six)
	of Weyl points in the first Brillouin zone.
	(d) Dispersion in the horizontal ($k_x-k_y$) plane (shown by blue colour in panel (a))
	containing the Weyl node near the $R$ point.
	Quasiparticles in this plane are similar to quasiparticles in graphene and are
	characterised by a non-trivial Berry phase of $\pi$.
	The arrows show the pseudospin $\langle\hbsigma\rangle$ (the Pauli matrices $\hsigma_i$ act in the space
	of the angular-momentum projections $J_z=0$ and $J_z=-1$).
	(e) Dispersion along the (yellow in panel (a)) plane consisting of vectors $\mathbf{k}=(\pi+{k_d}/{\sqrt{2}},\pi+ k_y,0.71\pi+{k_d}/{\sqrt{2}})$ containing the Weyl point.  Colour shows the weight of the  $\ket{1-1}$ state in the
	quasiparticle eigenstate, and arrows represent the pseudospin $\langle\hbsigma\rangle$.
	(f) Dispersion along the (green in panel (a)) vertical plane ($k_y-k_z$) containing the Weyl point near
	the $R$ point. For each momentum $\bk$ the colour represents  the weight of the $\ket{1-1}$ state .
	}
	\label{BZ}
\end{figure*}

{If an excitation with the angular momentum $J=1$ is created on site $i$, the dipole-dipole interaction
can transfer it to another site $j$, possibly changing the projection of the angular momentum on the
direction of the magnetic field; $|1\sigma\rangle_i\rightarrow|1\sigma^\prime\rangle_j$.
The quasiparticles in the system are thus hard-core bosons corresponding to the angular momentum-degrees
of freedom that hop from site to site as described by the effective Hamiltonian
(see Methods for a detailed derivation)
\begin{align}
	&\hat{H}_{\rm lat}=\sum_{i,j,\sigma,\sigma^\prime}M^{\sigma\sigma{'}}_{ij}\hat{b}^\dagger_{i\sigma} \hat{b}_{j\sigma{'}}=\sum_{\sigma\sigma{'}{\bf{k}}}M^{\sigma\sigma{'}}_{{\bf{k}}}\hat{b}^\dagger_{\sigma,{\bf{k}}} \hat{b}_{\sigma{'},{\bf{k}}},
	\label{HamLatt1}
	\\
	&M^{\sigma\sigma'}_{ij}=
	\langle 1\sigma|_{\mathbf{i}} \langle 00|_{\mathbf{j}}|\hat{H}_{\mathrm{dip}}\left(\hat{\mathbf{d}}_{\mathbf{i}},\hat{\mathbf{d}}_{\mathbf{j}},\mathbf{r}_{\mathbf{i}}-\mathbf{r}_{\mathbf{j}}\right)|00\rangle_{\mathbf{i}}|1\sigma'\rangle_{\mathbf{j}}
	\nonumber\\
	&-\sigma \delta_{\sigma\sigma'}\delta_{ij}B
	\label{HamLatt2} .
\end{align}
Due to the translational invariance, the single-excitation
Hamiltonian can be diagonalised in the basis of momentum states $\bk$,
with the results shown in Figs.~\ref{BZ}a,c.
In accordance with the above general phenomenological argument, for $B\neq0$ the dispersion has Weyl
nodes (six in the first Brillouin zone, Figs.~\ref{BZ}a,c).
}

To demonstrate the chiral nature of Weyl quasiparticles we show in Figs.~\ref{BZ}d,e,f
the pseudospins $\langle\hbsigma\rangle$
(with the Pauli matrices $\hsigma_i$ acting in the space of the $J_{\mathrm{z}}=-1$ and $J_{\mathrm{z}}=0$ angular-momentum projections)
for the eigenstates with momenta $\bk$ in the horizontal ($p_{\mathrm{z}}=0$), tilted $(p_{\mathrm{z}}=p_{\mathrm{x}})$, and vertical ($p_{\mathrm{x}}=0$) planes (Fig.~\ref{BZ}a)
that contain a Weyl node. Excitations in these planes are equivalent to quasiparticles in graphene, the
2D counterpart of a Weyl semimetal, and are characterised by the non-trivial Berry phase $\pi$.
Figs.~\ref{BZ}d,e,f demonstrate that the pseudospins $\hbsigma$ of these states are linked to their momenta
$\bp$, measured from the Weyl node, in agreement with the effective Hamiltonian~\eqref{WeylDisp}.

{\bf Effects of {quenched} disorder and dissipation.}
In general, quasiparticles in interacting systems have finite lifetimes due to
elastic and inelastic scattering processes. Indeed,
{deep optical lattices under consideration are ususlly not completely filled by particles
and thus inherently disordered due to the randomness of the particle disptribution.}
Also, spontaneous and dipolar collective emission from the internal $J=1$ levels to the ground state can lead to the decay of the excitations.

\begin{figure*}[ht!]
	\centering
	\includegraphics[width=.8\textwidth]{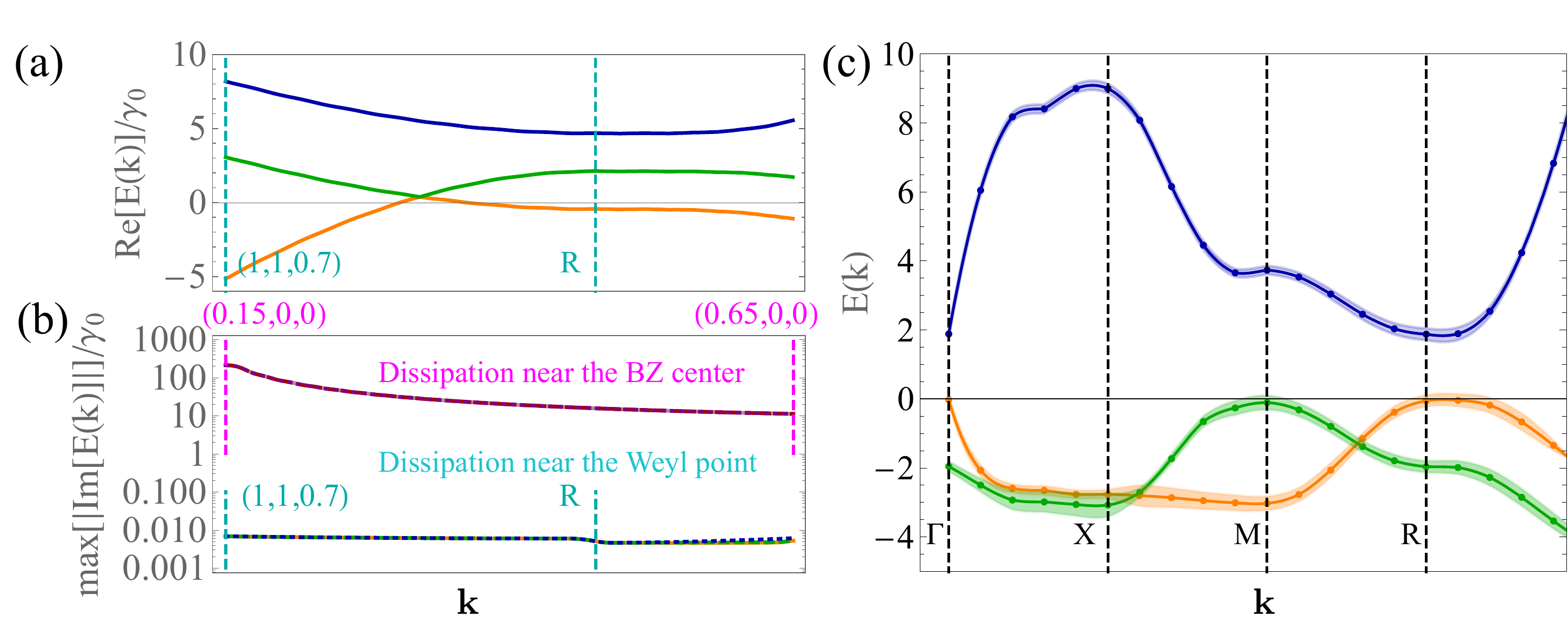}
	\caption{ {\bf Effects of disorder and dissipation on Weyl  quasiparticles} (a)  Real part of the quasiparticle dispersion in the presence of dissipation (including spontaneous emission and collective radiative decay). The parameters of the $J=1\leftrightarrow J=0$ transition correspond to those of
	the electronic $^3P_0-^3D_1$ levels of bosonic $^{88,84}\rm{Sr}$ atoms, trapped in a magic optical lattice potential with $a = 206.4$nm\cite{ShreckLesanovsky} at unit filling. Momentum $\bk$ is measured in units $\pi/a$.
	(b)  The upper bound on the inelastic scattering rate estimated from simulating the full quasiparticle spectra including all allowed elastic and inelastic dipolar processes. The
	dissipation is significantly suppressed  near the Weyl node (blue line). In striking contrast, the dissipation is enhanced close to the $\Gamma$ point (red line) due to the enhanced collective emission (superradiance).
	(c) Disordered case: Quasiparticle dispersion for a lattice with filling fraction $f = 0.93$ in the limit of small dissipation
	$k_0 a\ll1$ (the energy is measured in units $\gamma_0/(k_0 a)^3$).
	The line thickness shows the inelastic scattering rate.}
	\label{Fig2}
\end{figure*}

To analyse the effects of dissipation in a unit-filled lattice we compute numerically the quasiparticle dispersion
for retarded dipolar interactions, Eq.~(\ref{DipoleInteraction}), with parameters
of the $J=1$ to $J=0$ transition of the electronic ${}^3P_0- {}^3D_1$ levels of bosonic
${}^{88,84}\rm{Sr}$ atoms trapped in a magic optical lattice with $a=206.4$nm considered in
Ref.~\onlinecite{ShreckLesanovsky}. The wavelength and the dipole moment for this transition
are $2.6 \mu$m and $d=4.03$D, leading to the linewidth $\gamma_0= 290 \times 10^3$s$^{-1}$
and the dissipation parameter $ak_0\sim 0.5$.
Albeit quasiparticle damping in this regime is rather strong, it is significantly suppressed (by more than three orders of magnitude) near the Weyl nodes, as our simulations show, Fig.~\ref{Fig2}a,b.
Our results indicate that the quasiparticle scattering in such a system
would be dominated by quenched disorder rather than
by collective radiative decay or spontaneous emission.

To account for the effects of disorder we evaluate numerically the quasiparticle dispersion
for a lattice filling of $93\%$. This filling fraction could be achieved by preparing a cold bosonic Mott insulator using  moderate atom numbers that allow one to suppress doubly occupied states at the trap centre. Mott insulators
have already been realised with bosonic AEAs 
in the ground $^1{S}_0$ state \cite{Sugawa:review,Stellmer:review}. These atoms can be excited to the desired $^3P_0$ state by laser pulses\cite{Akatsuka:bosonMI}.

As our simulations demonstrate, the characteristic energy scales of Weyl excitations significantly
exceed the elastic scattering rate, demonstrating that the excitations could be conveniently observed in  current experiments.

\begin{figure*}[t]
\centering
\includegraphics[width=\textwidth]{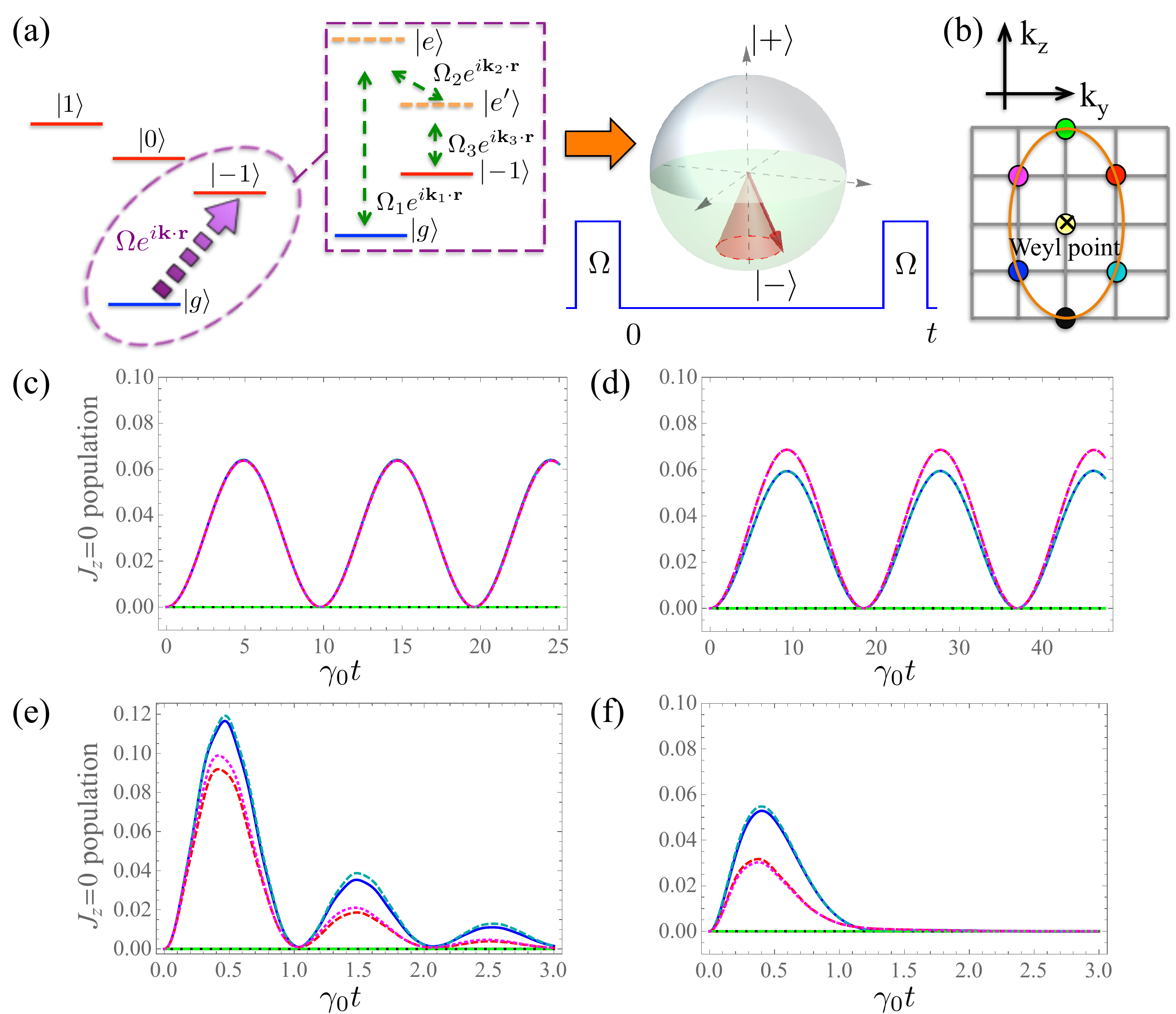}
  \caption{{\bf  Observation of Weyl quasiparticles}
  (a) Momentum-selective Ramsey spectroscopy:  interfering Raman beams create an excitation
with the angular momentum projection $J_{\rm z}=-1$ and with translational momentum $\bk$ near the Weyl node (see text). After a waiting time $t$ the second pulse is applied to measure the fraction of particles
in the $J_{\rm z}=0$ angular-momentum state.
(b) Six quasimomenta near the
Weyl node. (c) The fraction of particles in the $J_{\rm z}=0$ state oscillates as a function of time $t$
with the frequency $(E_\bk^1-E_\bk^2)/(2\pi)$, where $E_\bk^1-E_\bk^2$ is the energy splitting
between the two branches of the quasiparticle dispersion.
(d) The oscillations in the presence of dissipation for $k_0a\sim 0.5$.
(e) The oscillations in a $99\%$ randomly filled lattice.
(f) The oscillations in a $93\%$ filled lattice.  For (c) and (d) a 3D  cubic lattice of $100\times 100\times 100$ sites was used and we took advantage of the translational symmetry. For (e) and (f) a 3D cubic lattice of $10\times10\times10$ was used.}
  \label{fig:dyna}
\end{figure*}

{\bf Experimental observation.} For probing the Weyl character of the excitations we propose a
Ramsey protocol illustrated in Fig.~\ref{fig:dyna}a.  After preparing a Mott insulator of particles
in the $J=0$ state, a pulse of  interfering Raman beams is
used to create excitations in the $\ket{1,-1}$
angular-momentum state with translational momentum ${\bf k}$. Here we consider the case when ${\bf k}$ is set to be  close to the Weyl
point with intersecting $J_{\mathrm{z}}=0$ and $J_{\mathrm{z}}=-1$ branches. For the proposed  ${}^3P_0- {}^3D_1$ electronic levels in Sr, two intermediate states $|e\rangle, |e'\rangle$ could be used to  create the Raman pulses, imparting a net momentum to the atoms   proportional to ${\bf k}={\bf k}_1+ {\bf k}_2+{\bf k}_3$ (see Fig.~\ref{fig:dyna}a). A possible excitation level scheme consists in using  $5s6s$ ${}^3 S_1$ and  $5s6p$ ${}^3 P_1$ as the intermediate  $|e\rangle$ and $|e'\rangle$ levels respectively.
After a waiting time $t$, another pulse is applied to measure the fraction of particles in the $J_{\mathrm{z}}=0$
angular-momentum state.
Because of the interference of the two branches, this fraction oscillates
with the frequency $(E_\bk^1-E_\bk^2)/(2\pi)$, where the energy splitting $E_\bk^1-E_\bk^2$
between the two branches is linear in $\bk$ and vanishes near the Weyl node.

Another signature of the Weyl node is the strong dependency of the amplitude
of such oscillations on momentum $\bk$ near the node, as the amplitude
is determined by the  projection of the Bloch vector on $\hat{k}_{\mathrm{z}}$ (the magnetic field direction).


In Fig.~\ref{fig:dyna}(c-f) we show the fraction of particles in the $J_{\mathrm{z}}=0$ state as a function of time
at the end of the above described Ramsey protocol, for the six different quasi-momenta in the $k_{\mathrm{z}}-k_{\mathrm{y}}$ plane near the Weyl point  indicated in Fig.~\ref{fig:dyna}(b). Panel (c) shows the dynamics for an ideal unit filled lattice in the dissipationless limit $k_0a\ll1$.
Panel (d) shows the dynamics in the presence of dissipation for the experimentally relevant scenario discussed above.
The population dynamics in disordered systems is shown in
panels (e) and (f) for $99\%$ and $93\%$ filled lattices respectively. Quasiparticles scattering
on empty sites in a disordered system leads to the decay of the oscillations.


\section{Discussion} We demonstrated that Weyl quasiparticles generically emerge in 3D systems of polar particles in magnetic field. This opens intriguing prospects of observing chiral anomaly,
non-local electrodynamics, non-Anderson disorder-driven transitions, and other fascinating phenomena
in the realm of fully controllable atomic systems. We showed that observing Weyl excitations is currently possible in
arrays of AEA in 3D lattices, in particular, using the ${}^3P_0-{}^3D_1$ levels of bosonic $\rm{Sr}$ atoms.
Other experimentally convenient schemes, that deserve further exploration, include using
metastable levels of $\rm{Sr}$ or $\rm{Yb}$ atoms that can be trapped in magic lattices with spacings smaller than the wavelength\cite{Safronova2015}
or arrays of polar molecules with the rotational levels dressed to
avoid the splitting of $J=1$ levels in the presence of hyperfine interactions\cite{Yan2013}.
The long lifetimes and the topological character of Weyl excitations in interacting dipolar systems also
open new possibilities for implementing optical-lattice clocks with sensing capabilities beyond those
of non-interacting systems.


\section*{Methods}

{\bf Dispersion near Weyl nodes.} In this work, we define the quasiparticle dispersion as the poles
of the retarded Green's function averaged with respect to quenched disorder.

While long-wave quasiparticles ($\bk\rightarrow0$) are insensitive to the details
of the lattice potential, their effective Hamiltonian preserves rotation and inversion symmetries,
and in the absence of magnetic field-- time-reversal symmetry, with the generic form of the Hamiltonian
given by Eq.~\eqref{HamInit} and with the vector $\bomega$ parallel to the magnetic field.

We assume the existence of excitations with momentum $J=1$ and focus on the respective manifold of states
in  what follows. The dispersion of such excitations has three branches for each momentum $\bk$.

For momenta $\bk$ parallel to $\bomega$ the respective excitations have momentum projections
$J_{\mathrm{z}}=0$ and $J_{\mathrm{z}}=\pm 1$ on the direction $\bomega$. The branch with $J_{\mathrm{z}}=0$ intersects the branch
with $J_{\mathrm{z}}=\pm 1$ at momenta $\bK\parallel\omega$ such that
\begin{align}
	F(|\bK|,K^2,2)\pm\omega= F(|\bK|,0,2),
	\label{IntersectCond}
\end{align}
where we used that $\hat J^2=J(J+1)=2$ for the states under consideration.

The quasiparticle dispersion near the nodes can be found by expanding the Hamiltonian
in small momenta $\bp=\bk-\bK$.
Momentum deviation from a node along the $z$ axis leads to the splitting
$F\left[K+p_{\mathrm{z}},(K+p_{\mathrm{z}})^2,2\right]\pm\omega-F\left[K+p_{\mathrm{z}},0,2\right]$ between the intersecting branches.
Using that $[(\bK+\bp)\cdot\hat\bJ]^2\approx\frac{K^2}{2}+\frac{K^2}{2}\hsigma_{\mathrm{z}}+Kp_{\mathrm{z}}+Kp_{\mathrm{z}}\hsigma_{\mathrm{z}}\pm\frac{K}{\sqrt{2}}
(p_{\mathrm{x}}\hsigma_{\mathrm{x}}+p_{\mathrm{y}}\hsigma_{\mathrm{y}})$, with $\hsigma_i$ being the Pauli matrices in the space of momentum
projections $J_{\mathrm{z}}=+1$ $(J_{\mathrm{z}}=-1)$ and $J_{\mathrm{z}}=0$, we obtain the quasiparticle Hamiltonian~\eqref{WeylDisp}
with
\begin{align}
	\zeta(\bp) &=\frac{1}{2}F\left[K+p_{\mathrm{z}},(K+p_{\mathrm{z}})^2,2\right]-\frac{1}{2}F\left[K,K^2,2\right]
	\nonumber\\
	&+\frac{1}{2}F(K+p_{\mathrm{z}},0,2)-\frac{1}{2}F(K,0,2),\\
	v_\bot &=\pm\frac{1}{\sqrt{2}}K F^{(2)}(K,K^2,2)
	\label{Vperp}
	\\
	v_\parallel &=K F^{(2)}(K,K^2,2)
	\nonumber\\
	&+\frac{1}{2}[F^{(1)}(K,K^2,2)-F^{(1)}(K,0,2)]
\end{align}
where the upper (lower) sign in Eq.~\eqref{Vperp} applies to the intersection of the $J_{\mathrm{z}}=0$ branch
with $J_{\mathrm{z}}=+1$ ($J_{\mathrm{z}}=-1$), and
$F^{(1)}$ and $F^{(2)}$ are the derivatives of the function $F$ with respect to the first and the second argument.


{\bf Generic Hamiltonian of retarded dipole-dipole interactions.}
The dynamics of internal transitions $J=0\leftrightarrow J=1$ in a system of $N$ particles
is described by the Hamiltonian
\begin{align}
\hat{H}=\sum_{j=1}^N\Delta \hat{\bf c}_j^{\dagger}\cdot \hat{\bf c}_j+\sum_{{\bf q}\lambda}\omega_{\bf q} \hat{a}_{{\bf q},\lambda}^\dagger \hat{a}_{{\bf q},\lambda}
\nonumber\\
+d \sum_{j=1}^N {\bf E}({\bf r}_j)\cdot  (\hat{\bf c}_j^{\dagger}+\hat{\bf{c}}_j),
\end{align}
where the operator $\hat{c}_j^{\alpha\dagger}=|\alpha\rangle\langle 0|$ excites the $j$-th atom
from the ground state $|0\rangle$ to one of the Cartesian states $\alpha=x,y,z$ of the $J=1$ manifold
with energy $\Delta$; $d$ is the dipole moment of such a transition;
$\hat{a}_{{\bf q},\lambda}^\dagger$ and $\hat{a}_{{\bf q},\lambda}$ are the creation and annihilation
operators of a photon with momentum $\bq$, frequency $\omega$, and polarisation $\lambda$;
${\bf E}({\bf r})=i \sum_{{\bf q}\lambda}  \sqrt{\frac{\omega_{\bf q}}{2 \epsilon_0 V} }\hat{e}_{{\bf q},\lambda} (\hat{a}_{{\bf q},\lambda}^\dagger e^{-i \bf{q}\cdot {\bf r}}-\hat{a}_{{\bf q},\lambda}e^{i \bf{q}\cdot {\bf r}})$
is the operator of electric field, and $V$ is the volume of the system.

Eliminating the electromagnetic-field modes gives in the Born-Markov approximation the master
equation for for the density matrix of the particles\cite{GrossHaroche,James:emission}
\begin{align}
\dot{\hat{\rho}}=-i[\hat{H}_0+\hat{H}^{\mathrm{el}},\hat{\rho}] +\mathcal{D}(\hat{\rho}),
\label{Master}
\end{align}
where
$
\hat{H}_0=\sum_{j=1}^N\Delta \hat{\bf c}_j^{\dagger}\cdot \hat{\bf c}_j
$
is the Hamiltonian of the internal states of the particles,
and the effective interaction Hamiltonian is given by
\begin{align}
\hat{H}^{\mathrm{el}}_{j,k}=\sum_{j\neq k}\left(A(r_{jk})(\hat{\bf {r}}_{jk}\cdot  \hat\bd_j)(\hat{\bf {r}}_{jk} \cdot \hat\bd_k)+B(r_{jk})\hat\bd_j\cdot \hat\bd_k\right)
\end{align}
with $A(r)=\frac{3\gamma_0}{4d^2}\ty_2(k_0 r)$,
$B(r)=\frac{3\gamma_0}{4d^2}\sum_{n=0}^{1}(-1)^n \ty_n(k_0 r)/(k_0 r)$ and $\ty_n$ being the $n$-th-order
spherical Bessel function of the second kind, $k_0$ is the wavevector of the $J=0$ to $J=1$ transition,
$\hat{\bf d}_j^{\dagger} \equiv d \hat{\bf c}_j^{\dagger}$, $\mathbf{r}_{jk}=\mathbf{r}_j-\mathbf{r}_k, r_{jk}=|\mathbf{r}_{jk}|$, and $\hat{\mathbf{r}}_{jk}=\mathbf{r}_{jk}/r_{jk}$.

The operator $\mathcal{D}(\hat{\rho})$ in Eq.~\eqref{Master} accounts for dissipation and is given by
\begin{align}
\mathcal{D}(\hat{\rho})=-i \sum_{j, k}\left[ (\hat{H}^{\mathrm{di}}_{j,k} \hat{\rho}- \hat{\rho} \hat{H}^{\mathrm{di}*}_{j,k})+\Upsilon_{j,k}\right]
\end{align}
where $\hat{H}^{\mathrm{di}}_{j,k}={\mathcal A}(r_{jk})(\hat{\bf {r}}_{jk}\cdot  \hat\bd_j)(\hat{\bf {r}}_{jk} \cdot \hat\bd_k)+{\mathcal B}(r_{jk})\hat\bd_j\cdot \hat\bd_k$,
${\mathcal A}(r)= -i\frac{3\gamma_0}{4d^2}\tj_2(k_0 r)$, ${\mathcal B}(r)=-i\frac{3\gamma_0}{4d^2}\sum_{n=0}^{1}(-1)^n  \tj_n(k_0 r)]/(k_0 r)$, with $\tj_n $ being
the $n$-th-order spherical Bessel function of the  first  kind,
and $\Upsilon_{j,k}\propto
\hat{\bf c}_j\cdot \mathcal{T}_{j,k} \cdot \hat{\bf c}_k^{\dagger}$
is the so-called recycling operator\cite{GrossHaroche}
that does not affect the dynamics of a single excitation
and is thus omitted in the present paper.
Combining the interaction $\hat{H}^{\mathrm{el}}_{j,k}$ and dissipation $\hat{H}^{\mathrm{di}}_{j,k}$ terms we obtain
the effective (non-Hermitian) Hamiltonian~(\ref{DipoleInteraction}) of the dipole-dipole interactions.


{\bf Excitation dispersion in a deep lattice.}
As particles cannot move from site to site in a deep optical lattice,
the quasiparticles are represented by the angular-momentum degrees of freedom that
propagate through the system.
Assuming there is one particle per site and introducing bosonic operators $\hd_{i\sigma}^\dagger$ and $\hd_{i\sigma}$
for creating and annihilating the particle state $|1\sigma\rangle_i$ on site $i$ with angular momentum $J=1$
and projection $\sigma$ and the operators $\ha_{i}^\dagger$ and $\ha_{i}$
for creating and annihilating the momentum state $J=0$ on site $i$, the system Hamiltonian can
be rewritten as
\begin{align}
	\hat{H}=\quad&
	\nonumber\\
	\sum_{i,j,\sigma,\sigma^\prime}&\langle 1\sigma|_{i} \langle 00|_{j}\hat{H}_{\mathrm{dip}}\left(\hat{\mathbf{d}}_{i}
	\hat{\mathbf{d}}_{j},\mathbf{r}_{ij}\right)|00\rangle_{i}|1\sigma'\rangle_{j}
	\,\hd_{i\sigma}^\dagger\hd_{j\sigma^\prime}\ha_j^\dagger\ha_i
	\nonumber
	\\
	+\sum_{i,j,\sigma,\sigma^\prime}
	&
	\left[\langle 1\sigma|_{i} \langle 1\sigma^\prime|_{j}\hat{H}_{\mathrm{dip}}\left(\hat{\mathbf{d}}_{i}
	\hat{\mathbf{d}}_{j},\mathbf{r}_{ij}\right)|00\rangle_{i}|00\rangle_{j}
	\,\hd_{i\sigma}^\dagger\hd^\dagger_{j\sigma^\prime}\ha_j\ha_i\right.
	\nonumber
	\\
	&
	\left.
	+\langle 00|_{i} \langle 00|_{j}\hat{H}_{\mathrm{dip}}\left(\hat{\mathbf{d}}_{i}
	\hat{\mathbf{d}}_{j},\mathbf{r}_{ij}\right)|1\sigma\rangle_{i}|1\sigma^\prime\rangle_{j}
	\,\ha_j^\dagger\ha_i^\dagger\hd_{i\sigma}\hd_{j\sigma^\prime}\right]
	\nonumber
	\\
	-B \sum_{i\sigma}&
	\sigma \,\hd_{i\sigma}^\dagger\hd_{i\sigma}\ha_i\ha_i^\dagger
	+B_{\mathrm{J}} \sum_{i\sigma} \hd_{i\sigma}^\dagger\hd_{i\sigma}
	\nonumber
	\\
	+U\sum_{i\sigma}&
	\left[
	\hd_{i\sigma}^\dagger\hd_{i\sigma}\left(\hd_{i\sigma}^\dagger\hd_{i\sigma}-1\right)
	+\ha_{i}^\dagger\ha_{i}\left(\ha_{i}^\dagger\ha_{i}-1\right)
	\right.
	\nonumber
	\\
	&\left.
	+\hd_{i\sigma}^\dagger\hd_{i\sigma}\ha_{i}^\dagger\ha_{i}
	\right].
	\label{DeepLatticeMicro}
\end{align}
The first term in the Hamiltonian~(\ref{DeepLatticeMicro}) is responsible for moving angular-momentum excitations
from site to site; the angular-momentum state $|1\sigma\rangle_i$
can be transferred by the dipole-dipole interactions from site $i$ to another state $|1\sigma^\prime\rangle_j$
on site $j$ that initially was in the $J=0$ state.
The terms in the second sum in Eq.~(\ref{DeepLatticeMicro})
change pairs of sites $i$ and $j$ from the $J=0$ to $J=1$ angular-momentum states or vice versa.
The term $\propto B$ is the Zeeman energy.
The term $\propto B_{\mathrm{J}}$ accounts for the internal rotation (internal levels) of the particles.
The terms $\propto U=\infty$ in Eq.~(\ref{DeepLatticeMicro})
enforce the hard-core constraints for the bosons created by the
operators $\hd_{i\sigma}^\dagger$ and $\ha_i^\dagger$, taking into account that there is one
particle on each site.

In this paper we consider excitations on top of the ground state with all sites (particles) in the $J=0$
state. Exciting the internal degree of freedom of a particle on a site costs
the rotation energy $B_{\mathrm{J}}$ that significantly exceeds all the other energy scales,
except $U=\infty$, including the matrix elements $\sim|\hat{H}_{\mathrm{dip}}|$
of hopping of such angular-momentum degrees of freedom between sites
(for instance, for dipolar molecules and alkaline-earth atoms
$|\hat{H}_{\mathrm{dip}}|/B_{\mathrm{J}}\lesssim 10^{-6}$). As a result,
the number of sites excited to the $J=1$
states is conserved to a good approximation,
and the second sum in Eq.~(\ref{DeepLatticeMicro}),
that creates or annihilates pairs of $J=1$ excitations, can be neglected when considering the angular-momentum dynamics.

Therefore, the quasiparticles in the system are hard-core bosons that carry
angular-momentum ($J=1$) degrees of freedom and hop from site to site as described
by the effective Hamiltonian (\ref{HamLatt1})-(\ref{HamLatt2}) with
$\hb_{i\sigma}=\hd_{i\sigma}\ha_i^\dagger$.

{\bf Details of disorder averaging.}
Realistic systems of particles pinned in deep optical lattices are inherently disordered due to
the randomness of the spatial distribution of the particles. Each lattice site hosts either a
particle with probability $f$ or a vacancy with probability $1-f$.

For a small concentration of vacancies, excitations in the system are delocalised and
their dispersion is close to that in the disorder-free system but
acquires a small finite imaginary part $\text{Im} E_\bk$ due to the scattering on the vacancies.

In order to numerically obtain the quasiparticle spectra in such a disordered system
we diagonalise the Hamiltonian $\cH=\cH_0+\hV$, where $\cH_0$ is the excitation Hamiltonian
in the clean case and the operator $\hV$ models vacancies as sites with infinite potential $V(\br_i)=\infty$.
We compute the retarded Green's function
\begin{align}
G(\br_1\sigma_1,\br_2\sigma_2,E)
=
\sum_\alpha\frac{\psi^*_{\alpha\sigma_1}(\br_1)\psi_{\alpha\sigma_2}(\br_2)}{E-E_\alpha+i\eta}
\label{Gaveraged}
\end{align}
for multiple disorder realisations, where $\psi_\alpha$ and $E_\alpha$
are the eigenfunctions and eigenenergies for a particular disorder realisation,
$\sigma_1$ and $\sigma_2$ label projections of the angular momentum $J=1$,
and $\eta$ is a small positive number
introduced to ensure that the disorder-averaged Green's
function $\langle G(\br_1\sigma_1,\br_2\sigma_2,E)\rangle_{\text{dis}}$ is a smooth
function of its arguments for a given number of disorder realisations. At the same time, $\eta$
has to be chosen sufficiently small to not affect the results for the quasiparticle dispersion.
The energy $E$ has to be chosen close to the energies of the quasiparticles of interest.

Disorder averaging restores translational invariance, yielding
an averaged Green's function
that depends only on the coordinate difference $\br_2-\br_1$. Computing
the Fourier transform of the function $\langle G(\br_1\sigma_1,\br_2\sigma_2,E)\rangle_{\text{dis}}$
with respect to $\br_2-\br_1$ and diagonalising it in the angular-momentum space gives
$1/(E-E_{\bk n})$, where $n=1,2,3$ labels the dispersion branch for a given $\bk$,
$\text{Re}E_{\bk n}$ is the quasiparticle dispersion and
$-2\text{Im}E_{\bk n}$-- the scattering rate.

In this paper
we perform averaging over $100$ disorder realisations on a $10\times10\times10$ cubic
lattice with periodic boundary conditions
for the filling fraction $f=0.93$, close to that in the recent experiments\cite{kohl2005,schneider2008,esslinger2013}.
The results for the quasiparticle dispersion and scattering rates are shown in Fig.~\ref{Fig2}c.

The datasets generated in  the current study are available from the corresponding author on reasonable request.

\section*{Acknowledgments}
We thank M. Hermele and J. Ye for useful discussions and R. Nandkishore for feedback on the manuscript.  This work was supported by the NSF (PIF-1211914 and
PFC-1125844), AFOSR, AFOSR-MURI, NIST and ARO. MLW thanks the NRC postdoctoral fellowship programme
for support.  SVS has been also partially supported by the Alexander
von Humboldt Foundation through the Feodor Lynen Research Fellowship.

\section*{Author Contributions}
All authors contributed to all aspects of this work.

\section*{Additional information}

Competing financial interests: The authors declare no competing financial interests.

\end{document}